# 50 Years of Computational Complexity: Hao Wang and the Theory of Computation

**Nick Zhang**

Wuzhen Institute

nick@iwuzhen.org

**Abstract**

If Turing's groundbreaking paper [21] in 1936 laid the foundation of the theory of computation (ToC), it is no exaggeration to say that Cook's paper in 1971, "The complexity of theorem proving procedures" [4] has pioneered the study of computational complexity. So computational complexity, as an independent research field, is 50 years old now (2021) if we date from Cook's article. This year coincides with the 100th birthday of Cook's mentor Hao Wang, one of the most important logicians. This paper traces the origin of computational complexity, and meanwhile, tries to sort out the instrumental role that Wang played in the process.

1. **The origin of computational complexity**

The modern formation of the very concept of computational complexity can be attributed to Cobham, and Hartmanis and Stearns. Cobham's paper "The intrinsic computational difficulty of functions" and Hartmanis and Stearns's paper "On the computational complexity of algorithms" which gives the title of the field, were both written in 1963, but published later in 1964 [3] and 1965 [11], respectively. The both articles defined the time and space complexity of computation. Cobham even discussed using polynomial time as the measure of computational efficiency. In Harvard mathematics department, in order to obtain a PhD, in addition to a thesis, it requires a minor thesis and Cobham never bothered with a minor thesis. He entered the industry without finishing his PhD even though he already wrote his PhD thesis. He later helped Weslyan University to build its Computer Science Department, one of the earliest in the country. Hartmanis and Stearns received the 1993 ACM Turing Award. Hartmanis and Stearns's paper appeared in *Transactions of the American Mathematical Society*, but interestingly, Cobham's paper was first read in a conference of philosophy of science. Although the conference was closely related to logic and foundation of mathematics, it is clear that computational complexity was really new and homeless at that time.

Stephen Cook was a Ph.D. student at Harvard with Hao Wang as his advisor. Cobham was Wang's friend and Cook got to know Cobham and became familiar with his work. Hartmanis gave a speech on computational complexity at Harvard and Cook remembered it vividly. Cook was inspired by Cobham and trying to prove that multiplication requires $O(n^2)$ steps. Toom's work caused a big surprise [33]. Cook improved Toom's algorithm in his PhD Thesis [34], later referred as Toom-Cook algorithm. This algorithm is still of practical importance today. With the help of Stal Aanderaa [35], Cook showed that multiplication requires $O(n \log n/(\log \log n)^2)$ steps.

Prior to Cobham and Hartmanis and Stearns, Rabin, in 1959, suggested an axiomatic framework [38][39] that provided the basis for the abstract complexity theory developed by Blum [40]. The linear speed-up results from Hartmanis and Stearns were special cases of Blum's speed-up theorems.

A letter from Gödel to von Neumann on March 20th, 1956 was discovered by Harmanis in 1988 [12]. In the letter, Gödel pointed out that the difficulty of a problem can be expressed as a function of the steps needed to solve the problem on a Turing machine, and this function is the complexity of algorithm. Hence, the origination of theory of computational complexity was pushed back to as early as 1956. It is worth pointing out that Gödel realized the concept of the 'length of proof' in 1936 [42]. He put forward the idea of the number of steps instead of the number of symbols. But it was not until he saw the notion of Turing Machine that Gödel was convinced of Church-Turing's thesis. Gödel mentioned "It seems to me that this importance is largely due to the fact that with this concept one has for the first time succeeded in giving an absolute definition of an interesting epistemological notion, i.e., one not depending on the formalism chosen." [43]

In a batch of documents declassified by the National Security Agency (NSA) in 2012, one letter by John Nash to NSA in January 1955 appeared [15] [ see Figure 1]. It is handwritten on MIT stationery with eight pages in total. Apparently, Nash put forward the difference between polynomial complexity and exponential complexity in the previous year (1954), and he speculated that exponential complexity was useful for encryption algorithms. Nash and several colleagues at MIT at that time also discussed the so-called "exponential conjecture", among which Hoffman, later invented Hoffman code. Great minds think alike, Hoffman also had similar ideas. Michael Sipser, a leading theoretical computer scientist at MIT, believed Nash was the pioneer of the concept of complexity [18].

> MASSACHUSETTS INSTITUTE OF TECHNOLOGY
> CAMBRIDGE 39, MASS.
>
> DEPARTMENT OF MATHEMATICS
>
> Dear Major Grosjean,
>
> I have written RAND concerning the machine description. This was handwritten and was sent to NSA late last spring, I believe, or sent to someone there. Essentially the same machine description was once sent to a Navy communication center in Washington, I think.
>
> I have discussed the machine and general exponential conjecture with R.C. Blanchfield and A.M. Gleason who have worked for NSA.
>
> Recently a conversation with Prof. Hoffman here indicated that he has recently been working on a machine with similar objectives. Since he will be consulting for NSA I shall

Figure 1. Nash's letter to NSA

Even earlier, in 1953, Hao Wang wrote an article entitled "Recursiveness and calculability", in which he outlined the concept of speed function, which is actually the embryonic form of complexity. Unfortunately, the article has never been published. Hao Wang submitted it to the *British Journal of Philosophy of Science* in the summer of 1953. The review comments were very long. Hao Wang made a revision in 1954. After several times of communication with the editor, Hao Wang lost the patience of further revision and simply gave up. Until 1984, Wu Yunzeng, a logician from Peking University, saw the 1954 manuscript in Wang's office at Rockefeller University and suggested him include it in his anthology *Computation, Logic, Philosophy* [28]. Wang finally changed the title to "The Concept of Computability". I wonder if Mr. Wu had seen the communications between Wang and editors. Wang wrote an epilogue about this episode. In a way, we can accordingly say that the theory of computational complexity started in 1953.

I even traced the root the complexity theory to prosperous period of logic [31]. In 1939, Wittgenstein taught a course on "Foundations of Mathematics" in Cambridge. At this time, Turing had just returned from the Princeton to Cambridge and planned to teach mathematical logic. He happened to also name his own course "Foundation of Mathematics". When he learned that Wittgenstein was going to teach a course with the same name, he decided to attend Wittgenstein's class, which was regularly held at Wittgenstein's residence. Wittgenstein's lecture is Socratic, with no curriculum and handouts. More than 20 years after his death, his students gathered together the lecture notes and compiled them into a book *Wittgenstein's Lectures on the Foundations of Mathematics, Cambridge, 1939* [29]. The dialogue between Turing and Wittgenstein constitutes the large part. Turing's questions were challenging and Wittgenstein often times warded off. They had a profound discussion on the difficulty of proving a proposition, that is the length of proof.

| Time | | |
|---|---|---|
| 1939 | Turing, Wittgenstein | Difficulty and length of proof |
| 1953 | Wang | Speed functions |
| 1954 | Nash | Exponential conjecture |
| 1956 | Gödel | Number of steps for Turing machine to solve problems |
| 1959 | Robin | Degree of difficulty of computing a function |
| 1964 | Cobham | The intrinsic computational difficulty of functions |
| 1964 | Hartmanis, Stearns | On the computational complexity of algorithms |
| 1971 | Cook | NP-complete |

Table 1. Trace of computational complexity concept

In fact, in more widened sense, we can even push the origin of the theory of computation back to Charles Babbage and Ada Lovelace. Indeed, Babbage introduced "CARRYING THE TENS BY ANTICIPATION" [32] into Analytical Engine to improve efficiency of arithmetic operations of earlier Difference Engine. However, there must be clear and unquestionable milestones in the development of anything, and there are significant differences before and after the milestones. In this sense, Turing (1936/1937) and Cook (1971) have more significant and profound influences. Schmidhuber's recent cynical comments [41] about Turing shows his dissatisfaction with dominance of English-speaking academic community. But we have to look for the consensus about the significance of the results.

Cook raised the problem of P vs NP. This is one of the most important problems in both computer science and mathematics today. Cook won the Turing Award in 1982. P vs NP problem has also been listed as one of Millennium Prize Problems by the Clay Mathematics Institute. It is worth mentioning that Cook (1971) was published at the ACM SIGACT *Symposium on the Theory of Computing* (STOC), which later became one of the first-tier conferences in TOC. The importance of top-level conferences is even higher than that of magazines, which is a feature of rapidly developing computer science. The publication period of conferences is shorter than that of magazines.

In the early 1930s, the definition of "effectively calculable" was not as clear, just like the definition of "intelligence" now. In the early days, people preferred to use "calculable". Even Wang used "calculability" in 1953. By the 1960s, people generally used "computability". Church and Kleene predated Turing, in 1934, they knew that their λ-definable function is equivalent to the recursive function put forward by Herbrand-Gödel in 1933. They had formed Church Thesis, but Gödel didn't buy it until he saw Turing machine. He said Turing "It seems to me that this importance is largely due to the fact that with this concept one has for the first time succeeded in giving an absolute definition of an interesting epistemological notion, i.e., one not depending on the formalism chosen." Hence, the Church Thesis became Church–Turing Thesis.

## 2. Cook and P vs NP

Cook studied engineering as an undergraduate in Michigan, but switched to mathematics encouraged by a math professor. He went to Harvard and planned to study algebra. He got along well with Hao Wang who became his advisor. He got his PhD in 1966. He then went to Berkeley and stayed primarily in the Mathematics Department and also with the newly formed Computer Science Division from 1966 to 1970. He failed to get tenure and left in 1970 for University of Toronto. Richard Karp, another Turing Award winner and then a professor at Berkeley, recalled "It is to our everlasting shame that we were unable to persuade the math department to give him tenure." Cook later went to Toronto, not far from his hometown Buffalo, and worked as an associate professor both in the computer science and mathematics departments. At that time, the Computer Science Department was just set up to accept graduate students only. In the following year, his landmark paper was published, and he also got a tenure, working full-time in the more established CS Department. As sign of maturity the department began to accept undergraduates.

In the late 1960s, computer science just emerged as an independent discipline, and the mainstream of mathematics department did not take TOC seriously. There was a strong Logic group in the Berkeley Mathematics Department at the time, led by Tarski who once aspired to reserve one-tenth of the faculty positions of the Department to logic [31]. Logicians would understand ToC better, but it is obvious that not all logicians could see the importance of ToC, and in their eyes, ToC was just a half-hearted leftover of recursion theory, one of the four major branches of mathematical logic. Now CS Departments become logicians' dreamland. Hao Wang once said that the mathematical logic in Russell's time was 90% logic and 10% mathematics, but now it is 90% mathematics. Dijkstra, another Turing Award winner, was rejected by Amsterdam University for tenure at about the same time, and then moved to the United States to join the University of Texas at Austin.

Cook (1971) proved that the satisfiability problem (SAT) of propositional logic is NP-complete, which is the first NP-complete problem. There is a standard textbook presentation of the proof. The basic idea is: first, prove that SAT is in NP, which is very simple, as long as it can show that any SAT problem is polynomial verifiable. On the other hand, it is not straight forward that all NP problems can be reduced to SAT. Cook's method is to use propositional logic to describe all possible states of non-deterministic Turing Machine operated in polynomial time. It is precisely because Cook used propositional logic that the title of his thesis has the keyword "theorem proving".

It is interesting to note that Cook did not get the 'big' result in the initial draft to STOC. After his paper was accepted and he started writing the final version, it was then he came up the idea of completeness. STOC was fairly easy to get accepted that time. Cook credited Wang for giving him the idea [37].

Karp is the responsible for making Cook's theory to the popular. In 1972, he proved that 21 problems (most of which are combinatorial mathematics problems) can be mutually reducible with SAT [16]. This means that there is a large class of problems that have similar properties to SAT, that is, NP completeness. Many results of ToC are presented in a similar way, that is, to find the inherent commonness from seemingly irrelevant theories. Church and Kleene proved that λ-definable and recursive functions are equivalent, and proved that a large number of number-theoretic function are recursive.

Leslie Valiant, a Turing Award laureate, was shocked when he heard a recording of an IBM meeting in the office of his mentor Mike Patterson in 1972. The part he heard was Karp's article about reducibility. Later he went back to study Karp's article all night, and excitedly went to Patterson's office the next morning and asked, "Have you read this article?" Patterson asked, "Why are you so excited? Have you solved it?". Computational complexity has then become Valiant's lifelong academic pursuit [22]. Valiant's undergraduate background is mathematics and physics. He said that substance of computer science was not formed stably at that time, and the content of computer courses taught in schools often changed, but Karp's articles made him a transforming experience. This is not unique to him, the work of Cook and Karp was a beacon for a group of people standing at the intersection of mathematics, logic and computer science at that time. It is closely related to this that the research on "computer" became "Computer *Science*". Stephen Smale, a Fields Medal winner, was addicted to ToC in his

later years. He once said that "NP-complete problem is a gift from computer science to mathematics."

### 3. Hao Wang's Contribution to Theory of Computation

Wang received his Ph.D. in 1948 under the Harvard philosopher and logician Quine. At that time, his main interests were logic, foundations and philosophy of mathematics. In the 1950s, Wang had the idea of going back to China, so he turned his research interest to more practical and computer-related research, specifically, theory of computation and mechanical theorem proving. Wang's contributions to computer science can be summarized in a timetable, as shown in Table 2. The following sections respectively introduce some important and interesting ones.

| Time | |
|---|---|
| 1953-1954 | Speed functions |
| 1954-1957 | Wang Machine |
| 1957-1965 | Mechanical theorem proving |
| 1960-1961 | AEA, Wang Tiles |
| 1966 | Cook got PhD |
| 1974-1976 | Co-NP |

Table 2. Hao Wang's contribution to Theory of Computation

Wang returned to Harvard from Oxford in 1961 as "Chair Professor of Computing Theory". According to Cook's recollection, although Hao Wang was a professor of mathematical logic, his affiliation is not in the department of mathematics nor the department of philosophy where his mentor Quine was in, but in the department of applied physics. This recollection may not be accurate. Harvard was expanding its department of applied science and engineering at that time. In 1914, Harvard and MIT had a plan to merge Harvard's applied science department with MIT, but later the Massachusetts Court prohibited the merger, which made Harvard's applied science and engineering department lag behind their neighbor. The 1960s was the time when Harvard revived applied science and engineering.

A neglected fact is that the method used by Cook was inspired by Hao Wang. The core of this method is to use logic to describe a problem to be solved, thus define a problem class. Turing (1936/1937) used first-order logic to describe states of a Turing Machine, while Cook (1971) used propositional logic to describe polynomial time non-deterministic Turing Machine. Cook (1971) quoted the Wang (1962) [24], in which "Wang Tiles" was invented as an intermediate link between logic and the problem described. The brilliantness of Cook's work comes from that method, just like Godel's incompleteness theorem and Turing's halting problem. This encoding and reducibility technique is later widely used in complexity theory. Cook later credited Wang's inspiration [37]. Cook is familiar with Wang's thoughts and methods very well, and his conclusion on NP-completeness is very similar to Wang's. Only that Turing and Wang are talking about predicate calculus, and Cook about propositional calculus."

Hao Wang had the habit of compiling his papers into book forms. Wang (1962) is 33-pages long, and also appeared in his anthology *Computation, Logic, Philosophy* [28].

Cook and his student Reckhow's paper [6] in 1974 was the second most cited one among all his works. The SAT problem is to test whether a set of logic formulas have satisfiable truth value assignments. Theorem proving is to test unsatisfiability, that is, there are no satisfiable truth values. Theorem proving procedure is generally proof-by-contradiction, that is, unsatisfiability problem, which is the complement of NP problem, also known as "Co-NP problem". This article started proof complexity [7].

Wang also cooperated with Dunham, a veteran of IBM, to study the complexity of tautology. The results were published in the *Annals of Mathematical Logic* [8] in 1976. The tautology is the unsatisfiability problem put forward, and it is of course Co-NP problem too. Wang and Dunham are interested in the Co-NP problem mainly because it is closely related to theorem proving. The most commonly used resolution method in theorem proving has been proved to be not polynomial [10]. But it is unclear now whether NP and Co-NP are equivalent. This is one of the most talked over topics in the 2019 conference to commemorate Cook's 50-year teaching at the University of Toronto. The problem of prime factoring belongs to NP∩Co-NP, but it is unknown whether it is NP-Complete. If the prime factoring is NP-Complete, then NP=Co-NP. Many problems in NP∩Co-NP are polynomial, and the most well-known is linear programming. But we do not know if P=NP∩Co-NP. It is worth mentioning that the prime factoring is the most important link between quantum computing and traditional theoretical computer science.

Wang was intensively studying Marxism, particularly materialist dialectics at that time. It was rather rare to be distracted by the Co-NP problem with his old friends and for their early interest in theorem proving. Wang must also have had an insight into the profundity of Cook's work and be attracted by it.

A little-known anecdote is that Wang recommended Chinese mathematician Hong Jiawei to Cook as a visiting scholar. During his stay in Toronto, Hong put forward the once sensational Similarity Principle, that is, the cost of simulation among reasonably strong computing devices is polynomial [13], which means that there is no difference in principle between those computing devices. The Similarity Principle is actually a strong version of the Church-Turing Thesis. Just like the Church-Turing Thesis, it is combination of observation and rationality rather than a mathematical theorem or physics law. Hong Jiawei proved that almost all major computing models can simulate each other in polynomial cost. The Similarity Principle has not been mentioned much now, mainly because it has become the working assumption in Theory of Computation. If quantum computing can solve problems that cannot be effectively solved on Turing Machine in polynomial time, it might lead to the defeat of the strong Church-Turing Thesis. As a side note, at first glance, compared with computability problem, complexity problem seems to have less to do with philosophy, but as Scott Aaronson put, complexity actually has more philosophical bearing than computability [2].

## 4. An episode: Wang Machine

Von Neumann architecture has two core ideas, the stored program and random-access memory (RAM). The tape of Turing Machine is sequential rather than RAM. In most cases, people equate von Neumann architecture with the stored program, and takes RAM for granted. Von Neumann long pointed out that the stored program is essential Universal Turing Machine and the original idea should be attributed to Turing. Wang noticed the disconnection between theory and practice in computer design. His paper "A Variant to Turing's Theory of Computing Machines" [23] was first read at an ACM meeting in 1954 and was later reprinted in *JACM* in 1957, in which Wang simplified and enhanced the original Turing Machine. It made the computational model not only a thought experiment, but also more feasible in engineering. The model was called "Wang Machine". In Wang Machine, the finite automaton in Turing Machine was replaced by a series of program instructions. Programming on Wang machine is very similar to programming in an assembly language on a modern computer.

Marvin Minsky was a big fan of Hao Wang. He was not only the forerunner of AI, but also deeply involved in ToC. The only one of his students who received Turing Award is the theorist Manuel Blum, whose PhD thesis is about theory recursive function. Minsky published a monograph on Theory of Computation in 1967: *Computing: Finite and Infinite Machines* [14]. The value of this book is underestimated. The book was organized in a fashion that is precisely clear and easy to understand. It is still enjoyable and productive to read today. Of course, it should be more valuable than the notorious *Perceptron*, another book of his published two years later, which denigrated neural networks and almost killed the field. Minsky spoke highly of Wang Machine in the book, calling it "the first formulation of a Turing-machine theory in terms of computer-like models". This led to a batch of work to revise the original Turing Machine, and Theory of Computation in 1960s was full of papers on various machine models. Wang did not think highly on this line of research. He said: The authors was just to find an excuse to brag to their engineering colleagues that they are theorists, but then turned back and whispered to the real theorists that they were doing the practical work.

## 5. Another episode: Wang Tile

Hao Wang found some subclasses of first-order logic when doing research in mechanical theorem proving, and AEA was one of them. AEA is a formula with only three quantifiers in the prenex normal form, where A is the universal quantifier and E is the existential quantifier. AEA is proved to be undecidable. This is a bit like reducing the SAT problem to 3SAT, which is still NP-Complete. In order to solve and explain the AEA problem, Wang designed Wang Tile, which is a series of tiles to cover the floor. The four sides of a tile can have different colors, and the constraint condition is that the edges of two adjacent tile must have the same color. The general tiling problem is unsolvable. Robert Berger, another student of Wang, reduced Turing Machines to tiling problems so that any Turing Machine can be realized with Wang Tile. In the following work of Berger and others, various sub classes of AEA problems are further sorted out.

Wang Tile later found wider applications beyond logic and ToC. In 1965, Wang wrote a popular article in *Scientific American* to introduce Wang tiles [25]. Mathematicians and theoretical computer scientists started expanding Wang's results, among them are Raphael Robinson, Don Knuth, John Conway and so on. Later, Roger Penrose, who won the Nobel Prize in Physics, extending Wang Tile from quadrangle to polygon, and made a variant, Penrose Tile, which is useful in studying atomic arrangement in quasicrystals. Wang Tile became a tool of combinatorial mathematics, and later played an important role in proving the reversibility of two-dimensional cellular automata is undecidable. Interestingly, Berger, Wang's student, later joined MIT's Lincoln Lab in the department of research and development of integrated circuits. This was probably influenced by Wang Tile, and there were continuously new patents under his name till 2017.

## 6. Reflections and Conclusions

Wigderson, a theoretical computer scientist, a winner of Abel Prize, one of the most prestigious prizes in mathematics, talks about the relationship between ToC, especially complexity theory, and several other scientific branches [30], just like Feynman did in the beginning of *The Feynman Lectures on Physics*, where the relationship between physics and chemistry, biology, astronomy, geology and even psychology are discussed. Wigderson believes that ToC is useful for other disciplines. ToC should be a part of liberal education, just as mathematics and physics have been the core of liberal education since ancient Greece. If we quantitatively compare the Church-Turing Thesis with Newton's Laws, and ask which is more important and fundamental, there may be no standard answer. It is arguable Church-Turing Thesis is to digital society what Newtonian mechanics is to industrialized society, both should be part of everyone's education. They set boundaries and possibilities for other disciplines, which is commonly known as the "first principles".

Unlike Richard Feynman, Steven Weinberg, C.N. (Frank) Yang and other physicists who despised contemporary continental philosophy, Scott Aaronson, in his long article [2], has sincerely tried to persuade philosophers to learn some Theory of Computation, especially computational complexity. This is a bit like casting pearls before swine: understanding Theory of Computation requires some intellectual capability; as someone caustically put that practitioner of contemporary philosophy are those who want to write but lacking writing skills.

As a philosopher, Hao Wang has a solid mathematics background, and he has made profound contributions to the many branches of mathematical logic and ToC. At the same time, he was deeply involved in the important events of his time. It is a pity that he died too early and did not express his thoughts more systematically, just like Gödel. It may be because he did not find the appropriate utterance.

Hao Wang is a bridge between Theory of Computation and computational complexity. In fact, his involvement in complexity theory was earlier than his work in mechanical theorem proving. His contribution to complexity theory was somewhat neglected. There are external reasons for this: Turing (1936) and Cook (1971) are landmarks. On the other hand, there are internal reasons, as Wang turned to philosophy in his most productive years and then showed little interest in ToC in his later years. From 1986 to 1987, my mentor, Professor Yang Xueliang of University of Science and Technology of China, visited Wang at Rockefeller, and they had lunch together, Yang asked about the Co-NP and the scheduling problem, but Wang said that he was no longer interested in technical issues. The Continuum Hypothesis, after Gödel's work in the 1940s, had always been the secret project that mathematical logicians worked hard at home, just like all number theorists are secretly trying to prove Riemann Hypothesis. It is said that Wang also worked hard on Continuum Hypothesis at one time, but after the problem was solved by Paul Cohen in 1966, Wang stopped working on any big technical problems. Wang once compared his experiences with those of physicist Frank Yang, his roommate at college days. Yang left Princeton Institute of Advanced Studies and went to The State University of New York at Stony Brook, as if he entered the secular and earthly world, guided a large group of students. While Wang went to Rockefeller from Harvard, seemed to hide in the ivory tower. As Rockefeller has no undergraduate students, so he could spend more time doing philosophy with no need to teach. Rockefeller has a unique system. Every professor has a laboratory named after them. The Wang Lab used to have gathered the most important figures in logic at that time. But later, Rockefeller changed its direction to focus on biological sciences, and reduced the investment in mathematics, physics and others. The Wang Lab only spared its head, and he really became a world unto himself [31].

Robert Soare, a logician at the University of Chicago, asked himself "Why Turing and not Church" when commenting on Turing (1936) in 2013 [20]. His beautifully written piece made comparison with Donatello, an artist in the early Renaissance, and Michelangelo. Both of them had sculpture statues with David as the theme. But Michelangelo's *David* is one of the most exquisite works in art history. Donatello is more than 60 years older than Leonardo Da Vinci, while Michelangelo is 23 years younger than Leonardo. Soare quotes the comments of art historians and thinks that Michelangelo perfectly embodies the concept of the ideal artists in modern sense, even when compared with Leonardo. Soare metaphorically refers to Turing as Michelangelo in Theory of Computation. We can also ask "Why Cook, but not Hartmann, not Cobham, not Nash, not even Turing nor Wittgenstein" for analogy of computational complexity, simply because Cook is Michelangelo in complexity theory.

What about Hao Wang? He is more like Leonardo Da Vinci, one generation older than Michelangelo, more versatile and omnivorous. Charles Parsons, a Harvard philosopher who knew Wang well, wrote in the conference commemorating Gödel's 90th birthday "Hao Wang as a Philosopher" [17], and aptly compared Wang and Gödel, who regarded Wang as his confidant. Both of them struggled tension between more systematic philosophy and their own background in more specialized research, which Parsons called *A Logical Journey* which is just title of Hao Wang's unfinished book.

**Chronology of Hao Wang**

Wang very much liked listing chronological tables in his books. He once claimed that this is the Chinese way of doing things. The table provided here reflects my best knowledge.

1921. Wang was born on May 20, in Jinan, China.

1937. Started his logic journey. Studied the *Logic* book by Chin Yueh-lin, one of founders of the modern logic in China.

1938-1943. Mathematics major in the National Southwestern Associated University, which was essentially a temporary combination of three top Chinese universities, Tsinghua, Peking and Nankai due to Japanese invasion during WW-II. Acquainted with C.N. Yang, later a Nobel Laureate in Physics. They were roommates at college. Audited Wang Sin-Jun's logic class.

1942. Started to communicate with Quine.

1943-1945. Studied with Chin Yueh-lin and obtained Master degree in Philosophy from Qinghua University.

1946. Sailed to U.S. on October on a scholarship from Harvard. Met Quine on November and then studied Quine's *Mathematical Logic*.

1947. Wrote the PhD thesis, *An Economic Ontology for Classical Analysis*, with Quine as advisor

1948. Obtained the PhD.

1948-51. Junior Fellow of the Society of Fellows at Harvard. Spent the third year of his Junior Fellowship at Zurich with Bernays.

1951-53. Assistant Professor at Harvard.

1953-54. Research Engineer at Burroughs (now Unisys). Proposed Wang machine. Seriously considered going back to China to serve newly established red China. It is fortunate that he did not do so finally. As a matter of fact, many scientists and engineers abroad went back to China during that time and later a lot of them were sent to prison or labor camp.

1954-55. John Locke Lecture, and Lecturer in Philosophy at Oxford.

1956-1961. Reader in Philosophy of Mathematics at Oxford.

1957-1958. Wrote the famous theorem proving program.

1960. Took one year leave from Oxford to work at Bell Labs.

1961. Wrote to Quine asking for position at Harvard. Returned to Harvard as a chair of computation theory, Gordon McKay Professor of Mathematical Logic and Applied Mathematics.

1964. Outspoken for China. According to Quine, Wang had been persistently 'unhappy' for twenty some years, from his early student years in China until his departure to Rockefeller.

1967. Moved to Rockefeller University which is essentially a medical research institute. Assembled a group of top logicians including D.A. Martin, Saul Kripke and others.

1972. Wang was among the first to go back to China immediately after Nixon's 'opening' of China.

1973. Worked on Chinese character recognition on computers, which fascinated Gödel who mistook that Wang was working on computer recognition of human characters.

1974. *From Mathematics to Philosophy* was published. It revealed some unpublished ideas of Gödel and Wang's reflections. Wang later felt disappointed that this work did not gain more impact.

1976. Rockefeller slashed several non-medical programs, and the logic group was among them. Wang remained as the only logic and mathematics professor since then. Now Rockefeller has only medical research staff with several biologically-minded physicists.

1994. Finished the draft of *Conversations with Gödel: From Logic to Philosophy* as promised several years ago. The book was later re-titled as *A Logical Journey: From Gödel to Philosophy* and published by MIT press. Another book draft, *Power of the Mind: From Computation to Intuition* is essentially a collection of essays since 1990s with modifications to fit into a book style, was also finished yet never published.

1995. Wang died on May 13, just one week to his 74th birthday. This year we also experienced other losses in logic community, Church, Gandy, Kleene, Rasiowa and Robinson. Among them, Wang was the youngest.